\documentclass{bmcart} 

\usepackage{amsthm,amsmath}
\usepackage[utf8]{inputenc} 

\def\includegraphics{}

\startlocaldefs
\usepackage{amssymb}
\usepackage{algorithmic}
\usepackage{algorithm}
\usepackage{mathtools}
\mathtoolsset{showonlyrefs}
\usepackage{url}
\usepackage{multirow}

\newcommand{\EE}{\mathbb{E}}
\newcommand{\NN}{\mathbb{N}}
\newcommand{\RR}{\mathbb{R}}
\newcommand{\One}{\mathbf{1}}
\newcommand{\intd}{\, \text{d}}

\newcommand{\vm}[1]{\ensuremath{\mathbf{#1}}}
\newcommand{\vms}[1]{\ensuremath{\boldsymbol{#1}}}
\newcommand{\transpose}{\top}
\newcommand{\eexp}[1]{ \text{e}^{#1} }
\newcommand{\eps}{\varepsilon}

\newcommand{\comments}[1]{}
\DeclareMathOperator*{\argmax}{arg\,max}

\newcommand{\pdf}{\text{pdf}}
\newcommand{\up}{\vm p} 
\newcommand{\Nmc}{N_{\text{MC}}} 
\newcommand{\dom}{D} 
\newcommand{\SD}{\Omega_{\text{s}}} 
\newcommand{\IRPo}{T} 
\newcommand{\freq}{\omega} 

\newcommand\change[1]{{\color{black}#1}} 

\definecolor{darkpastelgreen}{rgb}{0.01, 0.75, 0.24}
\newcommand\new[1]{{\color{black}#1}} 

\newcommand\pic[1]{{\color{black}#1}} 

\newcommand\car[1]{{\color{black}#1}} 

\endlocaldefs

\begin{document}

\begin{frontmatter}

\begin{fmbox}
\dochead{Research}


\change{\title{A Blackbox Yield Estimation Workflow with Gaussian Process Regression 
		Applied to the Design of Electromagnetic Devices}}


\author[
   addressref={aff1,aff2},                   
   corref={aff1},                       
   email={mona.fuhrlaender@tu-darmstadt.de}   
]{\inits{M}\fnm{Mona} \snm{Fuhrländer}}
\author[
   addressref={aff1,aff2},
   email={sebastian.schoeps@tu-darmstadt.de}
]{\inits{S}\fnm{Sebastian} \snm{Schöps}}


\address[id=aff1]{
  \orgname{Computational Electromagnetics Group, Technische Universität Darmstadt}, 
  \street{Schlossgartenstr. 8},                     %
  \postcode{64289}                                
  \city{Darmstadt},                              
  \cny{Germany}                                    
}
\address[id=aff2]{%
  \orgname{Centre for Computational Engineering, Technische Universität Darmstadt},
  \street{Dolivostr. 15},
  \postcode{64293}
  \city{Darmstadt},
  \cny{Germany}
}



\end{fmbox}


\begin{abstractbox}

\begin{abstract} 
In this paper an efficient and reliable method for stochastic yield estimation is presented. Since one main challenge of uncertainty quantification is the computational feasibility, we propose a hybrid approach where most of the Monte Carlo sample points are evaluated with a surrogate model, and only a few sample points are reevaluated with the original high fidelity model. Gaussian Process Regression is a non-intrusive method which is used to build the surrogate model. Without many prerequisites, this gives us not only an approximation of the function value, but also an error indicator that we can use to decide whether a sample point should be reevaluated or not.
\change{For two benchmark problems, a dielectrical waveguide and a lowpass filter, the proposed methods outperform classic approaches.}
\end{abstract}




\end{abstractbox}
%

\end{frontmatter}


\section{Introduction}\label{intro}
In mass production of \change{electrical} devices, e.g. antennas or filters, often one has to deal with uncertainties in the manufacturing process. These uncertainties may lead to deviations in important parameters, e.g. geometrical or material parameters. Those may lead to rejections due to malfunctioning. In this context, the quantification of uncertainty and its impact plays an important role, also with regard to later optimization. According to Graeb~\cite[Chap. 2]{Graeb_2007aa} we define the yield as the percentage of realizations in a manufacturing process, which fulfills performance feature specifications. When dealing with electromagnetism, the performance feature specifications 
\change{are requirements involving partial differential equations (PDEs) describing the electromagnetic fields, i.e., Maxwell's equations. These can be solved numerically, e.g. with the finite element method (FEM).}
The most straightforward approach for yield estimation is the Monte Carlo (MC) analysis~\cite[Chap. 5]{Hammersley_1964aa}
\change{. The space of the uncertain parameters is sampled and the performance feature specifications are tested for each sample point. This} requires typically many evaluations of the underlying PDEs. Thus, the computational effort is one main challenge of yield estimation.

Over the last decades, various methods have been developed with the aim of reducing the computational effort of MC. 
One approach to achieve this is to reduce the number of sample points, e.g. by Importance Sampling (IS)~\cite[Chap. 5.4]{Hammersley_1964aa}. Another approach is to reduce the effort for each sample point, e.g. by using surrogate based approaches~\cite{Rao_1999aa,Rasmussen_2006aa,Babuska_2007aa}.
The cost for building most surrogate models increases rapidly with the number of uncertain parameters. Furthermore, there are counter examples where the yield estimator fails drastically, even though the surrogate model seems highly accurate, measured by classical norms or pointwise~\cite{Li_2010aa}. Therefore the same authors propose a hybrid approach. 
They focus their attention on critical sample points that are close to the limit state function, which is the limit between sample points fulfilling and not fulfilling the performance feature specifications.
Critical sample points are evaluated on the original high fidelity model, while the other sample points that are far from the limit state function are evaluated only on a surrogate model. 
\change{Because of this distinction, which leads to a combination of different model types, it is called a hybrid approach.}
Here, the choice of the surrogate model and the definition of \textit{close} and \textit{far} are crucial. 
In~\cite{Butler_2018aa}, a hybrid approach is proposed, using radial basis functions (RBF) for the surrogate model and an adjoint error indicator to choose the critical sample points. In~\cite{Fuhrlander_2020aa} a similar hybrid approach is proposed, using stochastic collocation with polynomial basis functions and also an adjoint error indicator. In this paper we combine these ideas and propose a hybrid approach using Gaussian Process Regression (GPR) for both, building the surrogate model and obtaining an error indicator in form of the prediction standard deviation given by the GPR. \change{The critical sample points are} used to improve the GPR model adaptively during the estimation process. \new{Further we investigate if sorting the sample points can increase the efficiency.}

Other research related to GPR based surrogate models for yield or failure probability estimation is conducted in~\cite{Bect_2012aa,Bect_2017aa,Xiao_2020aa,Zhang_2019aa}. 
\new{In~\cite{Bect_2012aa} various sorting strategies for GPR model training data are described and compared.}
In~\cite{Bect_2017aa}, the authors concentrate on the calculation of small failure probabilities with a limited number of function evaluations on the high fidelity model. They also use an adaptive GPR surrogate model, but do not combine it with a hybrid approach and therefore have no critical sample points that could be used to improve the GPR model. Instead, they distinguish between the sample points generated by Subset Simulation (Sequential MC) for error probability estimation and those generated as training data using a Stepwise Uncertainty Reduction technique to refine the GPR model adaptively.
In~\cite{Xiao_2020aa} and~\cite{Zhang_2019aa}, a GPR based surrogate model approach is combined with IS. Again, no hybrid approach is used. Adaptively, GPR model and IS density are improved by adding one or more sample points from the MC sample of the last iteration to the training data set, which are selected by a learning function and then calculated on the high fidelity model.
On the contrary, in practice it is often assumed that the design parameter deviations are small in a way that a linearization is valid~\cite[Online Help: Yield Analysis Overview]{CST2018}. This approach is obviously very efficient but it  is very difficult to determine on beforehand if the assumption is valid.

The paper is structured as follows. After setting up the problem, in Section 3 existing approaches and the concept of GPR are briefly described. Then the use of GPR for yield estimation, also in combination with a hybrid approach, is discussed. In Section 4, numerical results are presented using a benchmark problem, a simple waveguide, and a practical example, a low pass filter calculated with CST, before the paper is concluded in Section 5.

\section{Problem Setting}\label{setting}
\change{Even though the proposed ideas are generally applicable, in the following we will focus on problems from the electrical engineering where electromagnetic field simulations are necessary. This is the case, for example, when designing antennas or filters. Depending on the frequency, e.g. the electric field can be calculated to retrieve information about the performance of the device. Then the performance can be optimized by adapting the design.
In order to calculate the electric field on a simply connected bounded domain $\dom$, we start from Maxwell's formulation}
\begin{equation}
\nabla \times \left(\mu^{-1} \nabla \times \vm E_{\freq} \right)  - \freq^2 \eps \vm E_{\freq} = j \freq \vm{J} \quad \text{on~}\dom, 
\label{eq:E-field_strong_general}
\end{equation}
where $\vm E_{\freq}= \vm E_{\freq}(\vm x, \up)$ denotes the electric field phasor, $\omega$ the angular frequency, $\mu=\mu_r \mu_0$ the dispersive complex magnetic permeability, $\epsilon=\epsilon_r\epsilon_0$ the dispersive complex electric permittivity and and $\vm{J}=\vm{J}(\vm{x},\up)$ the phasor of current density. The vacuum and relative permeability are denoted by $\mu_0$ and $\mu_r=\mu_r(\vm{x},\up)$, the vacuum and relative permittivity respectively by $\epsilon_0$ and $\epsilon_r=\epsilon_r(\vm{x},\up)$.
Assuming suitable boundary conditions, building the weak formulation and discretizing with (high-order) Nédélec basis functions we derive the linear system
\begin{equation}
\vm A_{\freq} (\up) \, \vm e_{\freq}(\up) = \vm f_{\freq},
\label{eq:E-field_FEM}
\end{equation}
with system matrix $\vm A_{\freq} (\up)$, discrete solution $\vm e_{\freq}(\up)$, the discretized right-hand side $\vm f_{\freq}$, all depending on the design parameter $\up$ and the frequency $\freq$. For further details we refer to \cite{Jackson_1998aa,Nedelec_1980aa,monk2003finite}.
As quantity of interest (QoI)
\begin{equation}
Q_{\freq}(\up) = q \left( \vm e_{\freq}(\up) \right),
\label{eq:QoI}
\end{equation} 
we consider the scattering parameter (S-parameter), cf.~\change{\cite[Chap. 3]{Pozar_2011aa} and~\cite{Fuhrlander_2020aa}}, i.e., $Q_{\freq}(\up) := S_{\freq}(\up)$. In this case, $q$ is an affine linear function, \change{but this is no requirement for the following yield estimation methods.}

If there are uncertainties in the manufacturing process, the design parameters may be subject to random deviations. Therefore we model the uncertain parameter vector $\up$ as multidimensional random variable.
We assume $\up$ to be (truncated) Gaussian distributed (cf.~\cite{Cohen_2016aa}), i.e., $\up \sim {\mathcal{N_T}\left( \overline{\up}, \vms{\Sigma},\vm{lb} , \vm{ub}\right)}$ with mean value $\overline{\up}$, covariance matrix $\vms{\Sigma}$, lower and upper bounds $\vm{lb}$ and $\vm{ub}$ and \change{the corresponding probability density function~\cite{Wilhelm_2010aa} 
\begin{equation}
\pdf_{\mathcal{N_T}}(\up) = \begin{cases}
\frac{\eexp{-\frac{1}{2}(\up-\overline{\up})^{\transpose} \vms \Sigma^{-1} (\up-\overline{\up})}}
{\int_{\vm{lb}}^{\vm{ub}} \eexp{-\frac{1}{2}(\up-\overline{\up})^{\transpose} \vms \Sigma^{-1} (\up-\overline{\up})} \intd \up  } & \text{ if } \vm{lb} \leq \up \leq \vm{ub} \\
0 & \, \text{ else.}
\end{cases}
\end{equation}
%
Note that the definition of the yield and the proposed estimation method is independent of the chosen probability density function $\pdf(\up)$. The Gaussian distribution is a typical choice for modeling design parameters as uncertain. Here, we truncate it to avoid non-physical realizations of $\up$, e.g. negative distances.}
%
Following~\cite{Graeb_2007aa} we define the performance feature specifications as a restriction on our QoI in a specific interval, i.e.,
\begin{equation}
\left|S_{\freq}(\up)\right| \leq c \ \ \forall \freq \in  \IRPo_{\freq} = \left[\freq_{\text{l}},\freq_{\text{u}}\right] 
\text{ in GHz,}
\label{eq:pfs}
\end{equation}
where $c$ is a constant and $\freq$ 
\change{a range parameter from an interval $\IRPo_{\freq}$. Here, we identify $\freq$ with the frequency and $\IRPo_{\freq}$ with the frequency domain of interest.}
The safe domain is defined as the set containing all parameters, fulfilling the performance feature specifications, i.e.,
\begin{equation}
\SD := \left\lbrace  \up: \left|S_{\freq}(\up)\right| \leq c \ \ \forall \freq \in  \IRPo_{\freq} \right\rbrace.
\label{eq:SafeDomain}
\end{equation}
Then, the yield can be written as~\cite[Chap. 4.8.3, Eq. (137)]{Graeb_2007aa}
\begin{equation}
Y(\overline{\up}) := \EE [\One_{\SD}(\up)]
:= \int_{-\infty}^{\infty} \dots \int_{-\infty}^{\infty} \One_{\SD}(\up) \, \pdf(\up)  \intd \up,
\label{eq:Yield}
\end{equation}
where $\EE$ denotes the expected value and $\One_{\SD}$ the indicator function with value $1$ if the parameter $\up$ lies inside the safe domain and value $0$ otherwise.

\section{A GPR-Hybrid Approach for Yield Estimation}\label{sec:YieldEst}
In a MC analysis a large number of sample points is generated, according to the truncated normal distribution of the uncertain parameters, and evaluated in order to obtain the QoI. The fraction of sample points lying inside the safe domain is an estimator for the yield. Since the accuracy depends directly on the size of the sample, a classic MC analysis comes with high computational costs~\cite{Giles_2015aa}.
\change{In the past, various surrogate based approaches have been proposed. The idea is to approximate the QoI, i.e., 
find a mapping from the design parameter $\up$ to $\tilde{S}_{\freq}$, where $\tilde{S}_{\freq}$ is an approximation of $S_{\freq}$. This allows to evaluate the performance feature specifications~\eqref{eq:pfs} and thus the safe domain~\eqref{eq:SafeDomain} without solving a PDE for each sample point. 
}
The stochastic collocation hybrid approach proposed by~\cite{Fuhrlander_2020aa} showed that the computational effort can be reduced significantly while ensuring the same accuracy and robustness as with a classic MC method. 
Nevertheless, there are a few drawbacks.
First, since a polynomial collocation approach was used, the training data for the surrogate model must come from a tensorial grid and cannot be chosen arbitrarily. As a consequence the surrogate model cannot be updated easily, e.g. with the information from the evaluation of critical sample points. This could be handled by using regression,
but the second disadvantage would still remain: In order to distinguish between critical and non-critical sample points an adjoint error indicator was used. This requires the system matrices and the solution of the primal and the dual problem, which is not always given when using proprietary software
\pic{and can become very costly in case of non-linear QoIs.} The GPR-Hybrid approach we propose in this paper overcomes these issues.

\subsection{Gaussian Process Regression}\label{sec:GPR_basics}
Following Rasmussen and Williams~\cite[Chap. 2.2]{Rasmussen_2006aa}, the technique of Gaussian Process Regression can be divided into four mandatory steps and one optional step.
\paragraph{1. Prior: }
We make some prior assumptions about the functions we expect to observe. We write
\begin{equation}
(S_{\up})_{\up \in \vm P} \sim \mathcal{GP}\left( m(\up),k\left(\up,\up'\right) \right),
\end{equation}
if we expect the S-parameter to follow a Gaussian Process (GP) with specific mean $m$ and kernel function $k$. 
In the following we use the constant zero function as a starting value for the mean function.
\change{When the GP is trained, the mean value of the training data evaluations will be used. As kernel function we}
choose the squared exponential kernel, which is also known as RBF, i.e.,
\begin{equation}
k\left(\up,\up'\right) = \zeta \, \eexp{-\frac{\left|\up-\up'\right|^2}{2l^2}},
\end{equation}
with the two hyperparameters $\zeta \in \RR$ and $l>0$.
At this point we refer to Section~\ref{sec:Numerics} to see how we set the hyperparameters. For more information about hyperparameters in general, please refer to~\cite[Chap. 5]{Rasmussen_2006aa}.
\paragraph{2. Training data:} 
We collect data by evaluating sample points on the high fidelity FE model. The so-called training data set
\begin{equation}
\mathcal{T}=\left\lbrace \vm P=[\up_1,\dots,\up_n]^{\change{\transpose}}, \vm S = [S(\up_1),\dots,S(\up_n)]^{\change{\transpose}} \right\rbrace
\end{equation}
will be used to train the GP. 
\change{In the following, we generate these sample points according to the distribution of the uncertain parameters.}
\paragraph{3. Posterior:} 
In this step the information from the prior and the training data is combined in order to obtain a new GP, with updated mean and kernel function. We write
\begin{equation}
\vm K = \left[ \begin{array}{ccc} 
k(\up_1,\up_1) &  \dots & k(\up_1,\up_n)\\ 
\vdots & & \vdots\\
k(\up_n,\up_1) &  \dots & k(\up_n,\up_n)
\\\end{array} \right]
\ \text{ and } \
\vm m = \left[ \begin{array}{c}
m(\up_1) \\ \vdots \\ m(\up_n)
\\\end{array} \right],
\label{eq:K_m_Matrix}
\end{equation}
then the posterior distribution of the output $S_{\up}$ depending on the training data set $\mathcal{T}$ is given by
\begin{equation}
S_{\up}|\vm P, \vm S \sim \mathcal{N}(\vm m, \vm K).
\label{eq:posteriorDistr}
\end{equation}
\paragraph{4. Predictions:}
\change{For an arbitrary test data point $\up^{\star}$ 
the predicted distribution of the output $S_{\up^{\star}}$} depending on the training data set $\mathcal{T}$ and the test data point is given by
\begin{align}
S_{\up^{\star}}|\up^{\star}, \vm P, \vm S \sim \mathcal{N} ( &
m(\up^{\star}) + \vm k(\up^{\star},\vm P) \vm K^{-1} (\vm S - \vm m), \\
& k(\up^{\star},\up^{\star})-\vm k(\up^{\star},\vm P)\vm K^{-1} \vm k(\vm P,\up^{\star}) ),
\label{eq:predictedDistr}
\end{align}
with
\begin{align*}
&\vm k(\up^{\star},\vm P) &&\hspace*{-5.25cm}= \left[k(\up^{\star},\up_1),\dots,k(\up^{\star},\up_n)\right],\\
&\vm k(\vm P,\up^{\star}) &&\hspace*{-5.25cm}= \left[k(\up_1,\up^{\star}),\dots,k(\up_n,\up^{\star})\right]^{\transpose}.
\end{align*}
Thus, \change{GPR} predictions of the function value \change{$\tilde{S}_{\text{GPR}}(\up^{\star})$ and the standard deviation $\sigma_{\text{GPR}}(\up^{\star})$}
can be obtained. 
Please note, \change{that $\sigma_{\text{GPR}}(\up^{\star})$ is the standard deviation}
of the surrogate model and is not related to the design uncertainty\change{, i.e., $\vms \Sigma$}.
\paragraph{5. Model update (optional):}
A new data point $(\up_{\text{add.}}, S(\up_{\text{add.}}))$ can be used to update an existing GPR model. Therefore the training data set is updated to
\begin{equation}
\vm P_{\text{new}} = [\vm P, \up_{\text{add.}}]
\ \text{ and } \
\vm S_{\text{new}} = [\vm S, S(\up_{\text{add.}})],
\label{eq:TrainSet_new}
\end{equation}
as well as~\eqref{eq:K_m_Matrix} has to be updated according to
\begin{equation}
\vm K_{\text{new}} = \left[ \begin{array}{cc} 
\vm K &  \vm k(\vm P,\up_{\text{add.}})\\ 
\vm k(\up_{\text{add.}}, \vm P) & k(\up_{\text{add.}},\up_{\text{add.}})
\\\end{array} \right]
\ \text{ and } \
\vm m_{\text{new}} = \left[ \begin{array}{c}
\vm m \\ m(\up_{\text{add.}})
\\\end{array} \right].
\label{eq:K_m_Matrix_new}
\end{equation}
Then, predictions for a new test data point $\up^{\star}$ can be obtained by~\eqref{eq:predictedDistr} using the updated data from~\eqref{eq:TrainSet_new} and~\eqref{eq:K_m_Matrix_new}. 
\change{This update involves the factorization of the matrix $\vm K_{\text{new}}$ which is in the worst case of cubic complexity, i.e., $\mathcal{O}(n^3)$,  and 
can be reduced to $\mathcal{O}(r^2n)$ by using low-rank approximations, where $n$ is the number of training data points and $r$ the rank of the low-rank approximation~\cite[Chap. 8]{Rasmussen_2006aa}.
In conclusion, we assume that this effort is negligible in comparison to the high fidelity evaluations, i.e., solving~\eqref{eq:E-field_FEM}.}
For more detailed information about GPR we refer to~\cite[Chap. 2]{Rasmussen_2006aa}.

\subsection{Combining GPR and the Hybrid Approach}\label{sec:GPRHybrid}
The idea of the hybrid approach is saving computing time by evaluating most of the MC sample points on a cheap to evaluate surrogate model and only a small subset of the sample on the original high fidelity (e.g. FE) model. 
\change{The critical sample points, i.e., sample points close to the limit state function, are those which are evaluated on the high fidelity model.}
As mentioned before, the choice of the critical sample points is crucial, for efficiency and accuracy of this approach. In~\cite{Butler_2018aa} and \cite{Fuhrlander_2020aa} adjoint error indicators are used. Here, we take advantage of the GPR that provides an error indicator \change{in the} point $\up$ in the form of the \change{standard deviation $\sigma_{\text{GPR}}(\up)$}. 
The performance feature specification expects the inequality~\eqref{eq:pfs} to hold in the whole frequency interval $T_{\freq}$. However, we define a discrete subset $T_{\text{d}} \subset T_{\freq}$ and enforce only that the inequality holds for all $\freq_j \in T_{\text{d}}$. 
This means, for each frequency point $\freq_j$ a separate surrogate model is built, otherwise rational interpolation could be used, e.g.~\cite{Gustavsen_1999aa}. 
\change{Thus, the GPR model and the resulting prediction values and standard deviations depend on the frequency and are denoted by $\tilde{S}_{\text{GPR},\freq_j}(\up)$ and $\sigma_{\text{GPR},\freq_j}(\up)$ with $j=1,\dots,|T_{\text{d}}|$.}
\change{We apply a short circuit strategy, i.e., a sample point is not evaluated on the remaining frequency points if it has already been rejected for a previous one. 
This allows us to save computing time and does not affect the estimation result, except the case that a sample point has been rejected erroneously based on an underestimated standard deviation prediction.
}
\begin{algorithm}[t]
	\caption{Hybrid decision}
	\begin{algorithmic}[1]
		\STATE{\textbf{Input:} sample point $\up_i$, frequency range $T_{\text{d}}$, threshold $c$, safety factor $\gamma$, GPR surrogate models}
		\FOR{$j=1,\dots,\left|T_{\text{d}}\right|$}
		\STATE{Evaluate the GPR model and obtain $\tilde{S}_{\text{GPR},\freq_j}(\up_i)$ and $\sigma_{\text{GPR},\freq_j}(\up_i)$}
		\IF{$\left| \tilde{S}_{\text{GPR},\freq_j}(\up_i) \right| + \gamma \left|\sigma_{\text{GPR},\freq_j}(\up_i) \right|\leq c $}
		\change{
			\STATE{Performance feature specifications fulfilled for $\freq_j$}}
		\ELSIF{$\left| \tilde{S}_{\text{GPR},\freq_j}(\up_i) \right| - \gamma \left|\sigma_{\text{GPR},\freq_j}(\up_i) \right| > c $}
		\STATE{Classify $\up_i \notin \SD$ (not accepted) and stop}
		\ELSE
		\change{
			\STATE{$\up_i$ is a critical sample point}}
		\STATE{Evaluate the high fidelity model and obtain $S_{\freq_j}(\up_i)$\\
			\IF{$\left|S_{\freq_j}(\up_i) \right| \leq c $}
			\change{
				\STATE{Performance feature specifications fulfilled for $\freq_j$}}
			\ELSE
			\STATE{Classify $\up_i \notin \SD$ (not accepted) and stop}
			\ENDIF
		}
		\ENDIF
		\change{
			\IF{$j=\left|T_{\text{d}}\right|$}
			\STATE{Classify $\up_i \in \SD$ (accepted)}
			\ELSE
			\STATE{Continue with $j=j+1$}
			\ENDIF}
		\ENDFOR
	\end{algorithmic}
	\label{algo:HybridDecision}
\end{algorithm}
%
%
%
%
Further, we build separate surrogate models for the real part and the imaginary part of the S-parameter, and later combine them for the prediction. This guarantees (affin-)linearity of the QoI by avoiding the square root.
\change{Algorithm~\ref{algo:HybridDecision} shows the classification procedure for one sample point $\up_i$.}
Once the GPR models are constructed, a MC analysis is carried out on the surrogates. 
For each sample point a predicted S-parameter value and a predicted standard deviation are obtained. 
\change{Following the concept of sigma-levels~\cite{Kumar_2006aa}, the predicted standard deviation multiplied with a safety factor $\gamma$ is considered as an error indicator for the surrogate model. 
The value of $\gamma$ is problem dependent and can be derived by evaluating some test data points on the high fidelity model and on the GPR model and considering the ratio of the true error and the predicted error, i.e., the standard deviation.}
The predicted standard deviation multiplied with \change{the} safety factor $\gamma$ serves as a buffer zone.
If the performance feature specification~\eqref{eq:pfs} is (not) fulfilled for the predicted S-parameter value and all values in the range plus/minus this buffer zone the considered sample point is classified as (not) accepted, else it is classified as critical and reevaluated on the high fidelity model.
Then, the yield will be estimated by
\begin{equation}
\tilde{Y}(\overline{\up}) = \frac{1}{\Nmc} \sum_{i=1}^{\Nmc} \One_{\SD}(\up_i),
\end{equation}
where $\Nmc$ is the size of the MC sample.
A significant advantage of GPR is, that the model can be easily updated on the fly. 
Algorithm~\ref{algo:YieldEst} shows the process of yield estimation including updating the GPR models.
%
\begin{algorithm}[t]
	\caption{Yield Estimation with GPR}
	\begin{algorithmic}[1]
		\STATE{\textbf{Input:} initial GPR models for each frequency point $\freq_j \in T_{\text{d}}$, set of MC sample points $\up_i, i=1,\dots,\Nmc$, error tolerance $\varepsilon_t\change{\geq}0$, \change{batch size $N_{\text{B}}\in\NN$} }
		\FOR{$i=1,\dots,\Nmc$}
		\STATE{Classify $\up_i$ according to Algorithm~\ref{algo:HybridDecision} \label{AlgoStep:ClassifyStart}}
		\STATE{\change{Count number of online high fidelity evaluations $|\text{HF}_{\text{GPR-H}}^{\text{online}}|$}}
		\FOR{$j=1,\dots,\left|T_{\text{d}}\right|$}
		\STATE{Define $\mathcal{C}_j = \left\lbrace \up_i:\up_i \text{ classified as critical for } \freq_j \text{ in last } N_{\text{B}} \text{ MC evaluations} \right\rbrace$ \label{AlgoStep:Cj}}
		\ENDFOR \label{AlgoStep:ClassifyEnd}
		\IF{\change{$|\text{HF}_{\text{GPR-H}}^{\text{online}}|$ is an integer multiplier of $N_{\text{B}}$}}
		\FOR{$j=1,\dots,\left|T_{\text{d}}\right|$ \label{AlgoStep:UpdateStart}}
		\STATE{Initialize \change{$\varepsilon=\infty$}}
		\WHILE{$\varepsilon>\varepsilon_t$}
		\STATE{Set $\up_{\text{add.}} = \argmax_{\up_i \in \mathcal{C}_j} \left|\tilde{S}_{\text{GPR},\freq_j}(\up_i)-S_{\freq_j}(\up_i)\right|$  \label{AlgoStep:padd}}
		\STATE{Update GPR model for $\freq_j$ with sample point $\up_{\text{add.}}$}
		\STATE{Evaluate updated GPR model and obtain updated $\tilde{S}_{\text{GPR},\freq_j}(\up_i)$ for all $\up_i \in \mathcal{C}_j$}
		\STATE{Calculate $\varepsilon =  \max_{\up_i \in \mathcal{C}_j} \left|\tilde{S}_{\text{GPR},\freq_j}(\up_i)-S_{\freq_j}(\up_i)\right|$}
		\ENDWHILE
		\ENDFOR \label{AlgoStep:UpdateEnd}
		\ENDIF
		\ENDFOR
		\STATE{Estimate the yield with $Y(\overline{\up}) = \frac{\left|\SD \right|}{\Nmc}$}
	\end{algorithmic}
	\label{algo:YieldEst}
\end{algorithm}
Typically the computational effort of a surrogate based approach lies in the \textit{offline} evaluation of the training data.
Therefore we start with a small initial training data set. The resulting less accurate GPR model does not pose a problem in terms of yield estimation accuracy,
because the hybrid method still classifies all \change{MC} sample points correctly \change{as accepted or not accepted}.
\change{The only difference is, that there might be more critical sample points in the beginning, if the initial GPR surrogate has been built with a smaller training data set.}
Then, during the estimation process (\textit{online}), we use \change{critical} sample points to improve our GPR model. 
This update requires almost no additional computational effort, since these sample points were calculated in the hybrid method anyway. 
%
%
%
\change{In order to enable parallel computing even with model updates, we introduce so-called batches. Only after the calculation of $N_{\text{B}}$ high fidelity evaluations (possibly in parallel), a GPR model update is considered. With $N_{\text{B}}$ we refer to the size of the batches, setting $N_{\text{B}}=1$ indicates that no batches are used. If only a part of the critical sample points is added to the training data set for updating the GPR model, they}
are chosen in a greedy way: After \change{evaluating one batch of MC sample points,} 
the resulting critical sample points of the $j$-th frequency point are collected in the set $\mathcal{C}_j$ \change{(cf. line~\ref{AlgoStep:Cj} in Algorithm~\ref{algo:YieldEst})}. Then, the sample point for which the difference between the predicted value and the real value of the S-parameter is maximum will be included in the training data set \change{(cf. line~\ref{AlgoStep:padd} in Algorithm~\ref{algo:YieldEst})}. The GPR model is updated with the additional training data point and all sample points in $\mathcal{C}_j$ are evaluated on the new GPR surrogate model in order to obtain a new prediction. This procedure is repeated until the error is below a tolerance $\varepsilon_t$.
\change{Using the updated GPR model, the next MC sample points are evaluated until again $N_{\text{B}}$ sample points have been evaluated on the high fidelity model (in parallel) and GPR model updates are considered.
Without much extra cost it is also possible to reevaluate all already considered, non-critical sample points after each GPR model update.}
\change{At this point it can also be decided whether all critical sample points are added to the training data set or only a part. Especially when solving in batches, it can be advisable not to include all critical sample points in order to avoid adding to many, closely neighboring sample points.}
\change{If $\varepsilon_t=0$, all critical sample points are used to update the GPR model.}

\new{The proposed updating strategy can be modified by sorting the sample points with negligible costs. The idea is to start with the most promising sample points, i.e., those that contribute most to the improvement of the GPR model. This shall lead to more sample points classified correctly without being evaluated on the high fidelity model.
In~\cite{Bect_2012aa} different sorting criterions are described and compared. Here, we will focus on two criterions. The criterion proposed by Echard, Gayton and Lemaire in~\cite{Echard_2011aa}, which we will call the EGL criterion in the following, and a criterion based on our hybrid decision rule, which we will call the Hybrid criterion.}
%
\new{
The EGL criterion is given by
\begin{equation}
C_{\text{EGL}}(\up_i) := 
\min_{\freq_j} \frac{\left|\tilde{S}_{\text{GPR},\freq_j}(\up_i) - c \right|}{\left|\sigma_{\text{GPR},\freq_j}(\up_i)\right|},
\label{eq:EGLcriterion}
\end{equation}
where $c$ denotes the upper bound for the performance feature specification from~\eqref{eq:pfs}. Then, the sample points are sorted such that we start with the smallest value, i.e., $\min_{\up_i} C_{\text{EGL}}(\up_i)$~\cite{Echard_2011aa}. 
The Hybrid criterion is defined by
\begin{align}
C_{\text{H}}(\up_i) := 
\max_{\freq_j} &\left( c-(|\tilde{S}_{\text{GPR},\freq_j}(\up_i)| - \gamma\,|\sigma_{\text{GPR},\freq_j}(\up_i)|)\right) \notag \\
&\left( (|\tilde{S}_{\text{GPR},\freq_j}(\up_i)| + \gamma\,|\sigma_{\text{GPR},\freq_j}(\up_i)|) -c \right).
\label{eq:Hcriterion}
\end{align}
Per definition 
\begin{equation}
C_{\text{H}}(\up_i) \begin{cases}
> 0, &\text{ if } \up_i \text{ is critical}\\
\leq 0, & \text{ else}
\end{cases}
\end{equation}
holds. Using this criterion, the sample points are sorted such that we start with the largest value, i.e., $\max_{\up_i} C_{\text{H}}(\up_i)$. Algorithm~\ref{algo:Sorting} is a modification of Algorithm~\ref{algo:YieldEst} including the sorting strategy. Before the classification of each sample point is started, all sample points are evaluated on the GPR models and sorted according to the chosen sorting criterion, e.g. the EGL criterion or the Hybrid criterion. Nevertheless, in the sampling strategy proposed in Algorithm~\ref{algo:Sorting} the sorting criterion can be replaced by any other criterion. After updating the GPR model for one batch of MC sample points, the remaining MC sample points are reevaluated on the updated GPR model and sorted again, according to the chosen criterion. This is repeated until all sample points are classified.
}
\begin{algorithm}[t]
	\caption{\new{Sorting strategy for yield estimation}}
	\begin{algorithmic}[1]
		\new{
			\STATE{\textbf{Input:} initial GPR models for each frequency point $\freq_j \in T_{\text{d}}$, set of MC sample points $\up_i, i=1,\dots,\Nmc$, 
				batch size $N_{\text{B}}\in\NN$, sorting criterion }
			\STATE{Evaluate all sample points $\up_i$, $i=1,\dots,\Nmc$ on the GPR models}
			\STATE{Sort all sample points according to the chosen sorting criterion}
			\FOR{$i=1,\dots,\Nmc$}
			\STATE{Run lines \ref{AlgoStep:ClassifyStart}-\ref{AlgoStep:ClassifyEnd} from Algorithm~\ref{algo:YieldEst} (i.e., classify $\up_i$ and define $\mathcal{C}_j$)}
			\IF{$|\text{HF}_{\text{GPR-H}}^{\text{online}}|$ is an integer multiplier of $N_{\text{B}}$}
			\STATE{Run lines \ref{AlgoStep:UpdateStart}-\ref{AlgoStep:UpdateEnd} from Algorithm~\ref{algo:YieldEst} (i.e., update GPR models)}
			\STATE{Evaluate remaining sample points $\up_k$, $k=i+1,\dots,\Nmc$}
			\STATE{Sort the remaining sample points according to the chosen sorting criterion}
			\ENDIF
			\ENDFOR
			\STATE{Estimate the yield with $Y(\overline{\up}) = \frac{\left|\SD \right|}{\Nmc}$}
		}
	\end{algorithmic}
	\label{algo:Sorting}
\end{algorithm}

\section{Numerical Results}\label{sec:Numerics}
In the following we perform numerical tests on two examples, a dielectrical waveguide and a stripline low pass filter. The results of the waveguide are also compared with the estimates resulting from a linearization, which is common in industry.
\change{The computations have been carried out with the following configuration: Intel i7-8550U processor with four cores, $1.80$\,GHz and $16$ GB RAM. For solving the corresponding PDEs~\eqref{eq:E-field_FEM} with FEM, the frequency domain solver of CST Studio Suite\textregistered\ 2018~\cite{CST2018} has been used. The yield estimation has been carried out in python 3.7, using the scikit-learn package version 0.21.3~\cite{scikit-learn} for GPR.
Solving our simple models takes only about $15$ seconds in CST, while the factorization for the GPR model update is always $\ll1$ second.}

\subsection{Dielectrical Waveguide}\label{sec:Numerics_WG}
\begin{figure}[t]
	\includegraphics{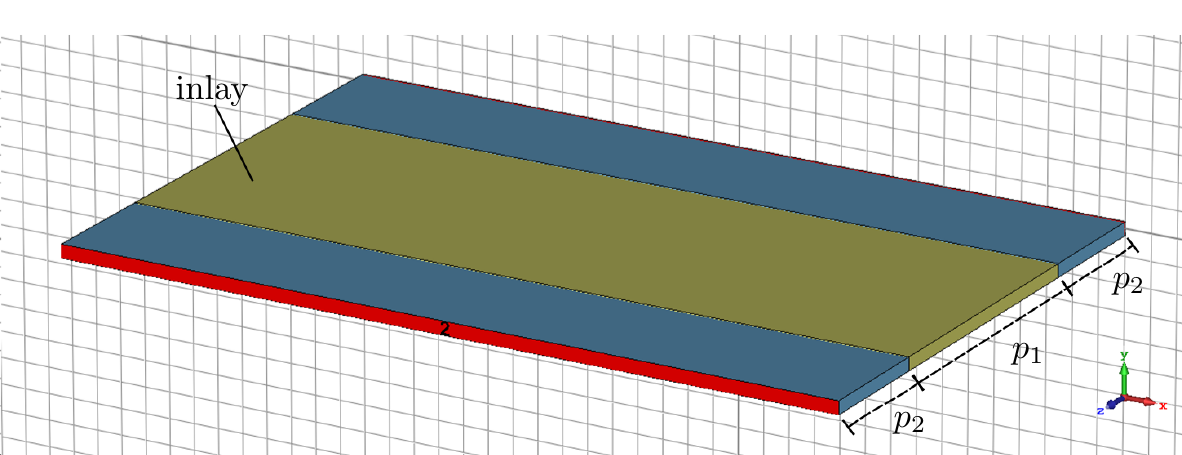}    
	\caption{Rectangular waveguide with dielectrical inlay of length $p1$ modelled in CST~\cite{CST2018}.}
	\label{fig:Waveguide}
\end{figure}

The benchmark problem, \change{an academic example}, on which we perform the numerical tests is a simple dielectrical waveguide, cf.~\cite{Fuhrlander_2020aa}. We consider two uncertain geometrical parameters, the length of the dielectrical inlay $p_1$, the length of the offset $p_2$ (see Figure~\ref{fig:Waveguide}), and two uncertain material parameters $p_3$ and $p_4$ with the following effect on the relative permeability and permittivity of the inlay
\begin{align*}
\epsilon_r &= 1+p_{3} + \left(1-p_{3}\right)\left(1+j\freq \left( 2 \pi 5 \cdot 10^9 \right)^{-1}\right)^{-1}, \\
\mu_r &= 1 + p_{4} + \left(2-p_{4}\right)\left(1+j\freq \left( 1.1\cdot 2\pi 20 \cdot 10^9 \right)^{-1}\right)^{-1}.
\end{align*}
The mean and covariance (in mm) is given by
\begin{equation}
\overline{\up} = [10.36, 4.76, 0.58, 0.64]^{\transpose} \ \text{ and } \ \vms{\Sigma} = \text{diag}\left([0.7^2, 0.7^2, 0.3^2, 0.3^2]\right).
\end{equation}
The distribution of the geometrical parameters is truncated on the left at $p_i-3$\,mm and on the right at $p_i+3$\,mm ($i=1,2$), the distribution of the material parameters is truncated on the left at $p_i-0.3$ and on the right at $p_i+0.3$ ($i=3,4$). The performance feature specifications are
\begin{equation}
\left| S_{\freq}(\up)\right|  \stackrel{!}{\leq} -24 \, \text{dB} \ \ \forall \freq \in \IRPo_{\freq} = \left[ 2 \pi 6.5, 2 \pi 7.5 \right] \text{ in GHz.}
\end{equation}
In this frequency range we consider eleven equidistant frequency points $\freq_j \in \IRPo_{\text{d}}$. 
\change{
A commonly used error indicator for MC estimation is given by~\cite{Giles_2015aa}
\begin{equation}
\tilde{\sigma}_{Y} := \sqrt{\frac{Y(\overline{\up})(1-Y(\overline{\up}))}{\Nmc}} \leq \frac{0.5}{\sqrt{\Nmc}},
\label{eq:stdYieldEst}
\end{equation}
where $\tilde{\sigma}_{Y}$ denotes the standard deviation of the yield estimator. Since the size of the yield is not known on beforehand, we estimate its upper bound by $Y(\overline{\up})=0.5$. We allow a standard deviation of $\tilde{\sigma}_{Y}=0.01$. According to~\eqref{eq:stdYieldEst} this leads to a sample size of $\Nmc = 2,500$. 
}
\change{Figure~\ref{fig:SampleSizeWG} shows values of MC yield estimators of the waveguide for different sample sizes. The black line indicates the most accurate solution we have calculated, i.e., $Y_{\text{MC}}$ for $\Nmc=10,000$. The gray shaded area indicates the $\tilde{\sigma}_Y$ level for the yield estimator of the corresponding sample size.
%
\begin{figure}[t]
	\includegraphics{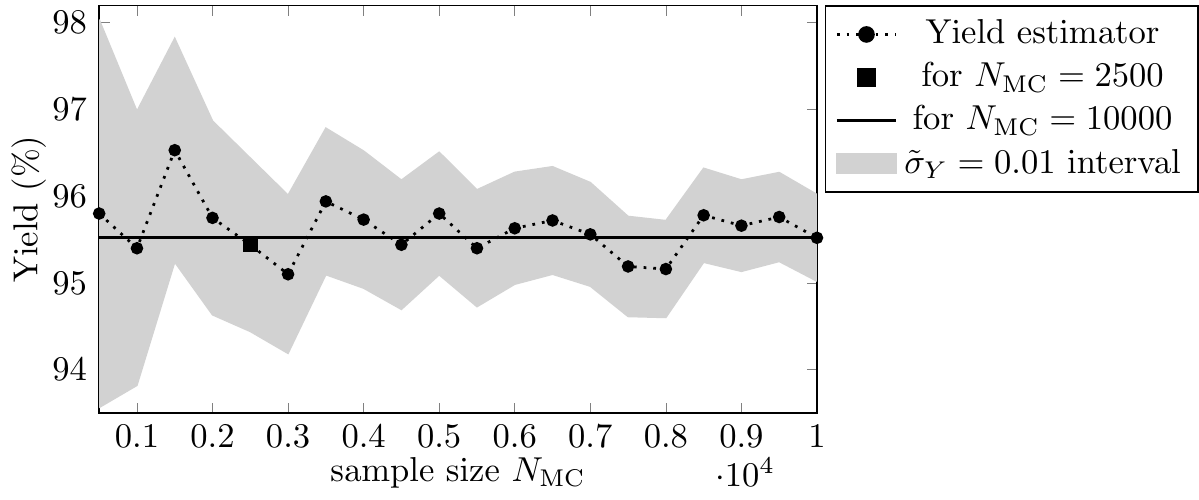}  
	\caption{\change{Values of MC yield estimators for different sample sizes $\Nmc$. The gray shaded area indicates the corresponding $\tilde{\sigma}_Y$ interval for each sample size.}}
	\label{fig:SampleSizeWG}
\end{figure}
}

%
\change{
The number of high fidelity evaluations before a possible update of the GPR model (batch) can be set to the number of parallel processors available, since these evaluations can be carried out in parallel. 
However, this value also has another effect:
a small number leads to more frequent model updates than a larger number. In general, more frequent model updates imply less critical sample points.} \new{We present tests with $N_{\text{B}}=50$, $N_{\text{B}}=20$ and $N_{\text{B}}=1$. The latter implies that no calculation in batches is used.
The error tolerance is set to $\varepsilon_t=0$, since this leads to the best results for the waveguide example, i.e., all critical sample points are added to the training data set.
Further we set the safety factor $\gamma=2$. This is a rather conservative choice, which may result in too many sample points classified as critical and evaluated on the high fidelity model. Thus, this may increase the computing effort, but it also leads to higher accuracy, since misclassification of sample points is avoided.  
}

\change{For the GPR, the} applied kernel is the product of a constant kernel representing $\zeta$ and \change{an RBF} kernel representing the exponential function with hyperparameter $l$. In scikit-learn, the hyperparameters have a starting point, in our case $\zeta_0 = 0.1$ and $l_0 = 1$, and then they are optimized within given bounds, in our case $b_{\zeta} = (10^{-5}, 10^{-1})$ and $b_{l}=(10^{-5}, 10^5)$, respectively.
\change{We allowed the hyperparameters to be tuned within $10$ iteration steps in order to find the most suitable values for our data. Due to this optimization, the initial setting does not affect the results of the yield estimation significantly. Further we}
set the noise parameter $\alpha = 10^{-5}$.
\change{This parameter defines the allowed deviation from the training data in the interpolation and is recommended to avoid numerical issues, e.g. due to mesh noise.}
\change{For more information about setting the hyperparameters we refer to~\cite[Chap. 2.3]{Rasmussen_2006aa} and~\cite{scikit-learn}.}
Once we have evaluated first training data points, the training data's mean is set as mean function of the GP.

For the simple waveguide a closed form solution of~\eqref{eq:E-field_FEM} exists, cf.~\cite{Loukrezis_WG}. However, we will refer to this solution as high fidelity solution in the following, since in practice a \change{computational} expensive FEM evaluation would be necessary at this point.
\change{
In the following we denote the number of high fidelity evaluations for a specific method with $|\text{HF}_{\text{method}}|$.}
\new{Further we introduce the number of effective evaluations $|\text{EE}_{\text{method}}| = \left\lceil \frac{|\text{HF}_{\text{method}}|}{N_{\text{B}}} \right\rceil$, which refers to the non-parallel high fidelity evaluations when using batches.}
The yield estimator with a pure, classic MC method serves as reference solution $\tilde{Y}_{\text{Ref.}} = 95.44\,\%$. 
\change{The number of high fidelity evaluations is $|\text{HF}_{\text{Ref.}}| = 26,360$. 
\new{Allowing parallel computing the number of effective evaluations would be $528$ for batch size $N_{\text{B}}=50$ and $1,318$ for batch size $N_{\text{B}}=20$.}
Please note, that the short circuit strategy mentioned in Section~\ref{sec:GPRHybrid} has been applied, i.e., a sample point is not tested for a frequency point if it has been rejected for a previous one. Without this short circuit strategy the number of high fidelity evaluations would be the product of the number of frequency points and the size of the MC sample, i.e., $|\IRPo_{\text{d}}| \cdot \Nmc = 11\cdot2,500=27,500$.} 
%

\change{In order to build the GPR models, an initial training data set is needed. It consists of random data points generated according the truncated Gaussian distribution ${\mathcal{N_T}\left( \overline{\up}, \vms{\Sigma},\vm{lb} , \vm{ub}\right)}$ of the uncertain parameters. The size of the initial training data set is chosen, such that the total costs, i.e., the sum of offline (initial training data) and online (critical sample points) costs, is minimal.} 
\new{Using batch size $N_{\text{B}}=50$, we tested different sizes of the initial training data set, see Table~\ref{table:InitialTrainSizeWG}. We proceed with the best performing number of ten training data points. For smaller initial training data sets the offline costs decrease, but the online costs increase. For larger initial training data sets it is the opposite.}
\begin{table}[t]
	\centering
	\begin{tabular}{l|ccc}
		& $|\mathcal{T}_{\text{I}}|=5$ & $|\mathcal{T}_{\text{I}}|=10$ & $|\mathcal{T}_{\text{I}}|=30$ \\
		\hline
		$|\text{HF}_{\text{GPR-H}}^{\text{offline}}|$ & $55$ & $\mathbf{110}$ & $330$\\
		$|\text{HF}_{\text{GPR-H}}^{\text{online}}|$ & \car{$306$} & \car{$\mathbf{226}$} & \car{$179$}\\
		$|\text{HF}_{\text{GPR-H}}^{\text{total}}|$ & \car{$361$} & \car{$\mathbf{336}$} & \car{$509$}\\
	\end{tabular}
	\caption{\new{Comparison of the number of high fidelity evaluations for the GPR-Hybrid approach $|\text{HF}_{\text{GPR-H}}|$ for different sizes of the initial training data set $|\mathcal{T}_{\text{I}}|$, using batch size $N_{\text{B}}=50$.}}
	\label{table:InitialTrainSizeWG}
\end{table}
%
Only this initial training data set is the same for all GPR models. Then, the estimation procedure with Algorithm~\ref{algo:YieldEst} is started. \change{After a batch of} $N_{\text{B}}$ critical sample points the GPR models are updated individually if there were critical sample points on them. 
\new{Table~\ref{table:UpdateStrategiesWG} shows the online high fidelity costs $|\text{HF}_{\text{GPR-H}}^{\text{online}}|$ and effective evaluations $|\text{EE}_{\text{GPR-H}}^{\text{online}}|$ for yield estimation with different updating strategies. In order to obtain the total costs, the costs for the initial training data set $|\text{HF}_{\text{GPR-H}}^{\text{offline}}|=110$, $|\text{EE}_{\text{GPR-H}}^{\text{offline}}|=\left\lceil \frac{110}{N_{\text{B}}}\right\rceil$ respectively, need to be added.
}
\begin{table}[t]
	\centering
	\begin{tabular}{l|cccccc}
		 & \multicolumn{2}{c}{no sorting} & \multicolumn{2}{c}{EGL criterion} & \multicolumn{2}{c}{Hybrid criterion} \\
		 & $|\text{HF}_{\text{GPR-H}}^{\text{online}}|$ & $|\text{EE}_{\text{GPR-H}}^{\text{online}}|$ & $|\text{HF}_{\text{GPR-H}}^{\text{online}}|$ & $|\text{EE}_{\text{GPR-H}}^{\text{online}}|$ & $|\text{HF}_{\text{GPR-H}}^{\text{online}}|$ & $|\text{EE}_{\text{GPR-H}}^{\text{online}}|$ \\
		\hline
		$N_{\text{B}}=1$&  $178$ & $178$ & \car{$146$} & \car{$146$} & $127$ & $127$\\
		$N_{\text{B}}=20$& \car{$197$} & \car{$10$} & \car{$163$}& \car{$9$}& $160$ & $8$\\
		$N_{\text{B}}=50$& \car{$226$} & $5$ & \car{$201$}& $5$& \car{$209$} & \car{$5$}\\
	\end{tabular}
	\caption{\new{Comparison of the number of online high fidelity evaluations  $|\text{HF}_{\text{GPR-H}}^{\text{online}}|$ and effective evaluations $|\text{EE}_{\text{GPR-H}}^{\text{online}}|$ for the GPR-Hybrid approach with different updating strategies.}}
	\label{table:UpdateStrategiesWG}
\end{table}
\new{In all cases, the} yield estimator is $\tilde{Y}_{\text{GPR-H}} = 95.44\,\%$, so we obtain the same accuracy as with pure MC. 
\new{With all updating strategies, the number of high fidelity evaluations can be reduced at least by factor \car{$78$}, in the best case by factor $111$, compared to classic MC.} 
%
\new{In the first setting, $N_{\text{B}}=1$, there are no batches (i.e., batches of size $1$). 
Without sorting and with both sorting criteria, this setting has the lowest number of high fidelity evaluations. However, parallel computing is not possible without batches, so the number of effective evaluations equals the number of high fidelity evaluations.
Using batches, 
the GPR models are not updated immediately, only after evaluating the complete batch. This leads to an increasement of the number of high fidelity evaluations. But, batches allow parallel computing (on $N_{\text{B}}$ parallel computers), i.e., the number of effective evaluations is much lower.
%


Further, we see that the number of high fidelity evaluations decreases when applying a sorting strategy. The GPR model is improved after the evaluation of a critical sample point. Due to the sorting, we start with the most critical sample points, so the GPR model improves fast and less sample points are categorized as critical. 
\car{The larger the batches are, the smaller the effect of sorting.}
}
%
\begin{figure}[t]
	\includegraphics{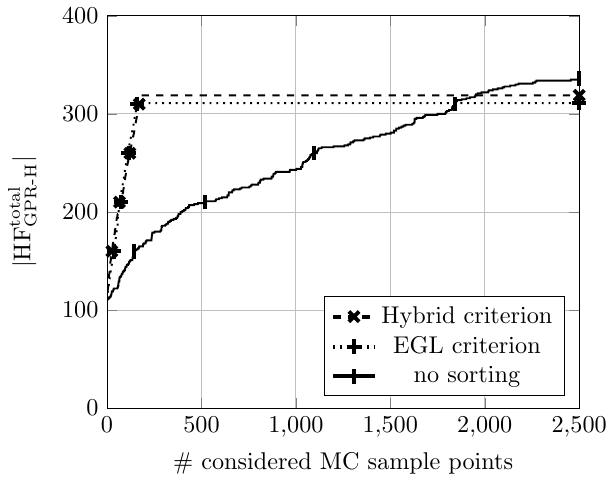}  %
	\caption{\pic{\car{Number of high fidelity evaluations  $|\text{HF}_{\text{GPR-H}}^{\text{total}}|$ over the number of considered MC sample points, using $N_{\text{B}}=50$ and different sorting strategies. The marks indicate the position where one batch is completed.}}}
	\label{fig:CriticalSamples_WG}
\end{figure}
\pic{Figure~\ref{fig:CriticalSamples_WG} shows the number of high fidelity evaluations  $|\text{HF}_{\text{GPR-H}}^{\text{total}}|$ over the number of MC sample points, which have been considered for classification. For the $0$-th considered MC sample point the offline costs are plotted, then the total costs. The different sorting strategies from Table~\ref{table:UpdateStrategiesWG} are compared for batch size $N_{\text{B}}=50$.
The marks indicate the position, where one batch is completed. We see, using sorting strategies, first all critical sample points are evaluated on the high fidelity model, then the non-critical sample points on the GPR model, i.e., the number of high fidelity evaluations increases early and then remains constant. The batches are filled within the first \car{$250$} MC sample points. Without sorting, the increasement is also a bit steeper in the beginning, but in general the batches are spread over the whole MC sample. In the end, the total number of high fidelity evaluations is \car{similar} for all strategies.}

\change{In the following, we compare these results to the results of the stochastic collocation hybrid approach proposed in~\cite{Fuhrlander_2020aa}. The hybrid method is the same, the difference lies in the choice of the surrogate model and the error indicator for defining the critical sample points. In~\cite{Fuhrlander_2020aa}, the surrogate model is built using an adaptive stochastic collocation approach with Leja nodes, 
\pic{which led to a maximum polynomial degree of three.}
Once the polynomial surrogate is built, it is not straightforward to update it during the estimation procedure. Thus, higher accuracy in the initial model is required. An adjoint error indicator is used to estimate the error of the surrogate model. Analogously to the standard deviation of the GPR model, this error indicator, multiplied with a safety factor.
Using this stochastic collocation hybrid approach, the same accuracy, i.e., the same yield estimator, was reached using
$|\text{HF}_{\text{SC-H}}^{\text{total}}| = |\text{HF}_{\text{SC-H}}^{\text{offline}}| + |\text{HF}_{\text{SC-H}}^{\text{online}}| = 330 + 165 = 495$ high fidelity evaluations. The number of training data points was chosen such that the method performs best.} 

\subsection{Comparison with a linearization approach}\label{sec:NumericsLin}
\begin{figure}[t]
	\includegraphics{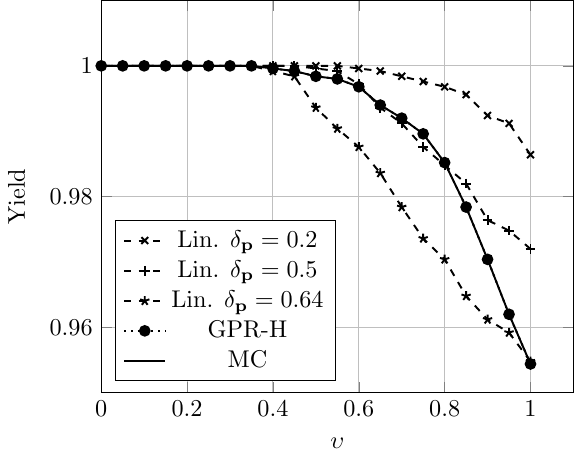}  
	\caption{Comparison of different yield estimation approaches over $\upsilon$: Reference solution is MC which coincides with the GPR-Hybrid approach. The linearization approach is plotted for different values of the step size $\delta_{\up}$.}
	\label{fig:Vgl_Lin_GPR}
\end{figure}

In practice, often a simple linearization of the QoI is used for the MC analysis, assuming that the design parameter deviations are small enough to obtain valid results~\cite{CST2018}. 
Therefore we compare the proposed GPR-Hybrid approach with linearization in the following. Linearizations means here, that we use a surrogate model, built by linear interpolation with two points in each dimension, i.e., in addition to $\up^0 = \left[p^0_1,p^0_2,p^0_3,p^0_4\right]^{\transpose}$ we consider the four nodes 
\begin{equation}
\up^k = \up^0 + \delta_{\up} \, \vm e_k, \ \ k=1,\dots,4,
\end{equation}
where $\vm e_k$ is the $k-$th unit vector and $\delta_{\up}>0$ the step size (if interpreted in the context of finite differences). Alternatively, derivative information could be used if available.
These five nodes are used to create a linear approximation according to
\begin{equation}
\tilde{S}_{\text{GPR},\freq_j}(\up^k) = 
\sum_{l=1}^{|\up^k|} \left(a_l \, \up^k_l \right) + a_{|\up^k|+1},
\end{equation}
where $|\up^k|$ is the length of the vector $\up^k$ and the $a_l$ are the coefficients of the linearization.
This model is setup for each frequency point $\freq_j$ and for the real and the imaginary part of the S-parameter separately. Then a MC analysis on the linear surrogate models is performed.
In Figure~\ref{fig:Vgl_Lin_GPR} we see the results of the yield estimation for different values of $\delta_{\up}$. 
We compare this to the MC solution on the high fidelity model as reference solution and the GPR-Hybrid solution from Section~\ref{sec:Numerics_WG}.
\change{
	We introduce $\upsilon\in [0,1]$ as a measure of the magnitude of deviation. The covariance matrix $\vms \Sigma$ is multiplied with this factor $\upsilon$ in order to obtain problem settings with varying magnitude of uncertainty, i.e., we consider  $\up \sim {\mathcal{N_T}\left( \overline{\up}, \upsilon \, \vms{\Sigma},\vm{lb} , \vm{ub}\right)}$ with different values for $\upsilon$.} 
%
For $\upsilon = 1$ we obtain the results of Section~\ref{sec:Numerics_WG}, for $\upsilon<1$ the scaled variance decreases and the yield estimator increases until for $\upsilon=0$ there is no uncertainty at all and the yield is $Y=1$ since $\overline{\up}$ is in the safe domain.
While the GPR-Hybrid solution exactly matches the reference solution \change{for all magnitudes $\upsilon$ of uncertainty}, we observe considerable deviations in the linearization model for any value of $\delta_{\up}$ (for $\upsilon>0.5$). These deviations decrease as expected with decreasing variance.

\subsection{Lowpass Filter}\label{sec:NumericsCST}
\begin{figure}[t]%
	\includegraphics[width=0.9\textwidth]{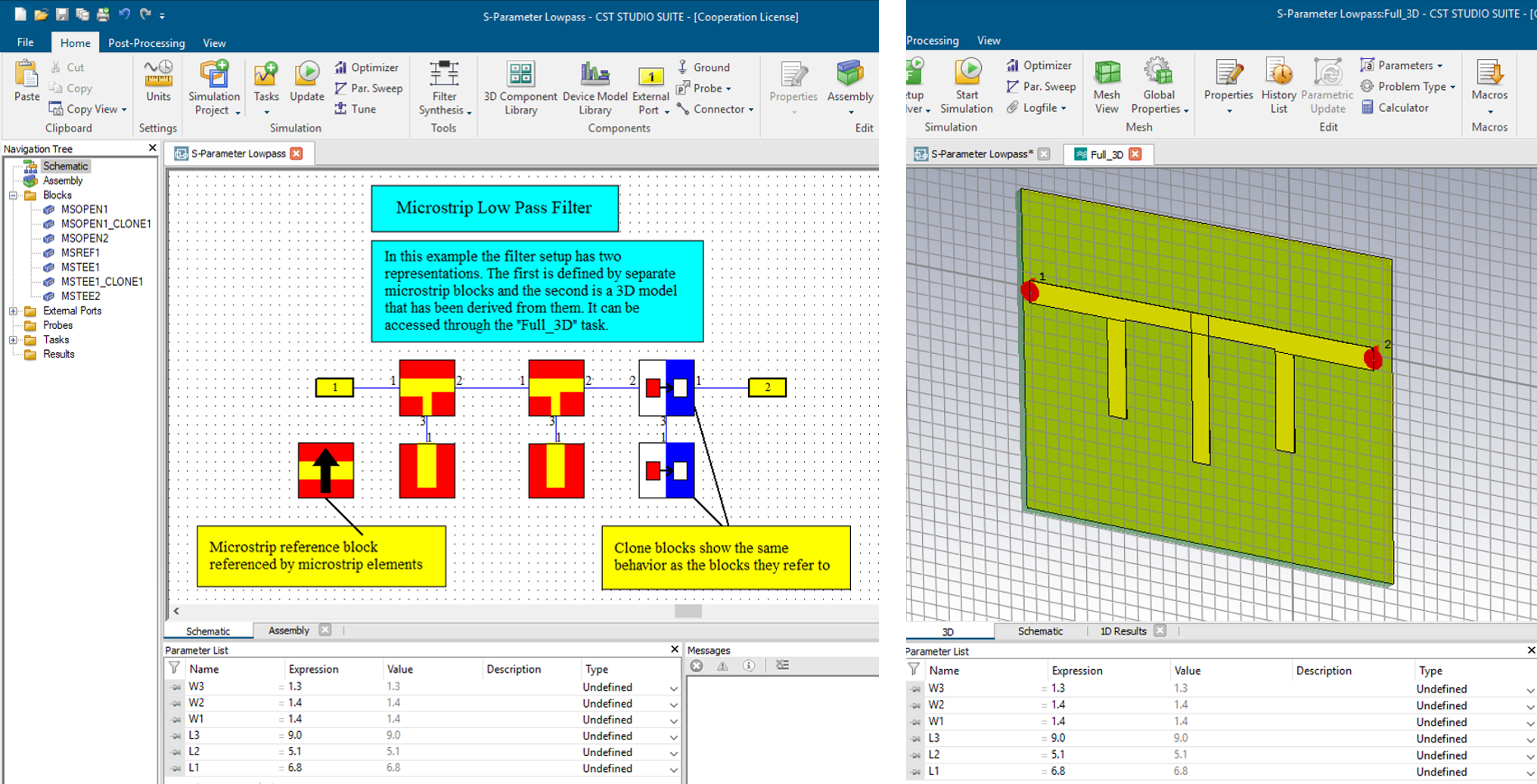} 
	\caption{Lowpass filter from examples library of CST Studio Suite\textregistered~\cite{CST2018}.}%
	\label{fig:CST_screenshot}%
\end{figure}

We consider as industrial example a stripline lowpass filter, see Figure~\ref{fig:CST_screenshot}, taken from the examples library of CST Studio Suite\textregistered~\cite{CST2018}. We consider six uncertain geometrical parameters $\vm g = [L_1,L_2,L_3,W_1,W_2,W_3]^{\transpose}$ describing length and width of the single blocks.
Again, we assume the uncertain parameters to follow a truncated Gaussian distribution with mean and covariance (in mm) given by
\begin{equation}
\overline{\vm g} = [6.8, 5.1, 9.0, 1.4, 1.4, 1.3]^{\transpose} \ \text{ and } \ \vms{\Sigma_{\vm g}} = \text{diag}\left([0.3^2, 0.3^2, 0.3^2, 0.1^2, 0.1^2, 0.1^2]\right).
\end{equation}
The distribution of $L_1$, $L_2$ and $L_3$ is truncated at $L_i\pm 3\,$mm ($i=1,2,3$), the distribution of $W_1$, $W_2$ and $W_3$ at $W_i\pm 0.3\,$mm ($i=1,2,3$). Since the requirement for a low pass filter is to allow low frequency signals to pass through while filtering out high frequency signals, in this example we have two performance feature specifications given by
\begin{align*}
\text{I.) } \ \left| S_{\freq}(\vm g)\right|  &\stackrel{!}{\geq} -1 \, \text{dB} \ \ &&\hspace{-3.1cm}\forall \freq \in \IRPo_{\freq,1} = \left[ 2 \pi 0, 2 \pi 4 \right] \text{ in GHz,}\\
\text{II.) } \ \left| S_{\freq}(\vm g)\right|  &\stackrel{!}{\leq} -20 \, \text{dB} \ \ &&\hspace{-3.1cm} \forall \freq \in \IRPo_{\freq,2} = \left[ 2 \pi 5, 2 \pi 7 \right] \text{ in GHz.}
\end{align*}
\change{As in the previous example we set $\tilde{\sigma}_Y=0.01$ which leads to a sample size of $\Nmc = 2,500$, according to~\eqref{eq:stdYieldEst}.}
\new{Again, we show test results for $N_{\text{B}}=50$, $N_{\text{B}}=20$ and $N_{\text{B}}=1$ and $\varepsilon_t=0$.} 
\change{Also, the kernel function and the hyperparameter settings are as in the previous example. The safety is set to $\gamma=3$. }
Further we consider eight equidistant frequency points $\freq_j \in \IRPo_{\text{d}}$, i.e., eight GPR surrogate models are built. The evaluation of the high fidelity model is implemented \change{in CST, using the default parameters of the frequency domain solver.}
The mathematical model is described in~\cite{Eller_2017aa}. An evaluation within CST calculates the S-parameter in a whole frequency range, i.e., for all considered frequency points $\freq_j \in \IRPo_{\text{d}}$. 
Therefore, with respect to this example, we look at the number of CST calls \change{$|\text{CC}_{\text{method}}|$} as a measure for the computational effort. \new{As before, in order to measure the efficiency for parallel computing, we introduce the number of effective calls $|\text{EC}_{\text{method}}| = \left\lceil \frac{|\text{CC}_{\text{method}}|}{N_{\text{B}}} \right\rceil$ as the number of non-parallelizable CST calls.}
\change{As reference value we consider the yield estimation with a pure Monte Carlo analysis. There, the computational effort is given by $|\text{CC}_{\text{Ref.}}| = 2,500$ and the estimated yield is $\tilde{Y}_{\text{Ref.}} = 87.08\,\%$.} 
\change{Again, the size of the initial training data set in the GPR-Hybrid approach has been chosen such that the total costs are minimal. This leads to an}
initial training data set of $|\mathcal{T}_{\text{I}}|=30$ sample points. This means we have an offline cost of \change{$|\text{CC}_{\text{GPR-H}}^{\text{offline}}|=30$}, because all frequency points are evaluated simultaneously in CST. 
Now we evaluate the $\Nmc$ sample points on the GPR model. If one sample point for one frequency point turns out to be a critical sample point, we evaluate this sample point for all frequency points with CST and use this information also for a possible update of the GPR models. 
\begin{table}[t]
	\centering
	\begin{tabular}{l|cccccc}
		 & \multicolumn{2}{c}{no sorting} & \multicolumn{2}{c}{EGL criterion} & \multicolumn{2}{c}{Hybrid criterion} \\
		 & $|\text{CC}_{\text{GPR-H}}^{\text{online}}|$ & $|\text{EC}_{\text{GPR-H}}^{\text{online}}|$ & $|\text{CC}_{\text{GPR-H}}^{\text{online}}|$ & $|\text{EC}_{\text{GPR-H}}^{\text{online}}|$ & $|\text{CC}_{\text{GPR-H}}^{\text{online}}|$ & $|\text{EC}_{\text{GPR-H}}^{\text{online}}|$ \\
		\hline
		$N_{\text{B}}=1$&  $320$ & $320$ & \car{$269$} & \car{$269$}& \car{$254$} & \car{$254$}\\
		$N_{\text{B}}=20$ & $352$ & $18$ & \car{$299$}& \car{$15$}& \car{$288$} & \car{$15$}\\
		$N_{\text{B}}=50$ &  $378$ & $8$ & \car{$339$}& \car{$7$}& \car{$326$} & \car{$7$}\\
	\end{tabular}
	\caption{\new{Comparison of the number of online CST calls  $|\text{CC}_{\text{GPR-H}}^{\text{online}}|$ and effective calls $|\text{EC}_{\text{GPR-H}}^{\text{online}}|$ for the GPR-Hybrid approach with different updating strategies.}}
	\label{table:UpdateStrategiesFilter}
\end{table}

\new{Table~\ref{table:UpdateStrategiesFilter} shows the online CST calls $|\text{CC}_{\text{GPR-H}}^{\text{online}}|$ and effective calls $|\text{EC}_{\text{GPR-H}}^{\text{online}}|$ for different updating settings.
Compared to classic MC analysis the computational effort can be reduced by a factor between $6$ and \car{almost $9$}, depending on the settings, while maintaining the accuracy.}
The lower savings in computational effort compared to the previous example of the waveguide is due to the fact that it is a more complex example on the one hand, but on the other hand also due to the simultaneous evaluation of all frequency points, because often a sample point is not critical for all frequency points.
\new{The results regarding the impact of the batch size $N_{\text{B}}$ remains similar as in the previous example. Without using batches, the lowest number of CST calls is needed. The number increase with the size of the batch, while the costs for effective calls decrease using parallel computations.
\car{Also, a slight improvement of efficiency by sorting the sample points could be observed.}
}

\section{Conclusions}\label{sec:conclusions}
A hybrid approach combining the efficiency of surrogate based approaches and the reliability and accuracy of the classic Monte Carlo method has been proposed. As surrogate model Gaussian Process Regression has been introduced and its standard deviation estimator was used as error indicator. Numerical results show that the computational effort can be significantly reduced while maintaining accuracy. 
This allows yield estimation in a reasonable time without the need for high performance computers as it would be the case with a pure Monte Carlo analysis.
Future research will focus on embedding the presented yield estimation methods in yield optimization. Furthermore, \change{interpolation in the direction of the range parameter} 
could be investigated.



\begin{backmatter}

\section*{Acknowledgements}
The  work  of  Mona Fuhrländer  is  supported  by  the Excellence  Initiative  of  the  German  Federal  and  State  Governments and the Graduate School of Computational Engineering at TU Darmstadt.
The authors would like to thank Frank Mosler of Dassault Syst{\`{e}}mes Deutschland GmbH for the fruitful discussions regarding the setup of the industrial example.
\new{Further, the authors thank Julien Bect of CentraleSupélec for very interesting discussions on generating and sorting training data for GPR models.}

\end{backmatter}
\end{document}